# Preliminary Design of CSNS-II Linac SRF LLRF

Zhexin Xie[1,2,3], Kai Guo[1,2,3], Zhencheng Mu[1,2,3], Xinpeng Ma[1,2,3], Nan Gan[1,2,3], Maliang. Wan[1,2,3], Bo Wang[1,2], Linyan. Rong[1,2,3], Hui Zhang[1,2], Hexin. Wang[1,2]

1. Institute of High Energy Physics(IHEP), Chinese Academy of Sciences (CAS),
Beijing 100049, China
2. Spallation Neutron Source Science Center (SNSSC), Dongguan 523803, China
3. University of Chinese Academy of Sciences, Beijing 100049, China

Keyword: LLRF, Superconductor, microTCA, Linac

*Abstract*

China Spallation Neutron Source(CSNS) target power will upgrade to 500 kW(CSNS-II) from 300kW, energy gain of H-Linac will up to 300 MeV from 80 MeV using about 50 superconductor cavities. LLRF is an important device for controlling the amplitude and phase of the SRF cavity field to be less than ± 0.3% and ± 0.3 °. The parameters and requirements for CSNS-II LINAC LLRF are presented here. The preliminary design work and algorithm verification progress and results at C-ADS Injector-I are introduced.

## 1. INTRODUCE OF CSNS-II AND SRF LLRF

The CSNS-II will increase target power from 100kW to 500kW. An SRF accelerator section will be added after the normal conductor section of the linear accelerator, which will increase the energy of the linear accelerator from 80MeV to 300MeV [1].

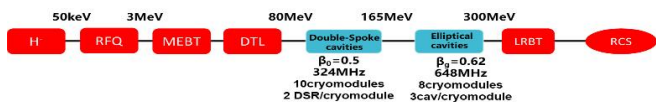

Figure 1: Schematic diagram of CSNS-II LINAC

There are two types of cavities in the SRF linear accelerator section, one is 324 MHz Double-spoke cavity, and the other is 648MHz elliptical cavity. There are about 50 SRF cavities totally. Both types of cavities are being processed and tested.

The specifications of the two types of cavities are shown in Table 1 based on simulation. The gradient of improved D-Spoke increases for 7.3 to 9 Mv/m ,while also having lower (LFD) Lorentz force detuning coefficient. A lower LDF will save RF power and improve the LLRF performance for pulse mode operation.

The Spoke cavity use a solid state amplifier ( SSA) with a 300kW peak RF power, and the elliptical use the domestic klystron amplifier with more than 600Kw peak RF power.

Because there is a RCS( Rapid circle synchronic ) after the Linac ,so the energy have to maintain at a high stability .The amplitude and phase stability should better that ±0.3% and ±0.3°. Parameter of SRF LLRF for CSNS-II LINAC is shown in table 2.

## 2. ALGORITHM VERIFICATION AT C-ADS INJECTOR I

Because there is not SRF accelerator at CSNS now, so we have algorithm verification at C-ADS Injector I.

C-ADS Injector-I includes one RFQ, two bunchers, and 14 spoke 012 cavities, all of which have frequencies of 325MHz. The Schematic of C-ADS Injector I is shown in Fig.2.

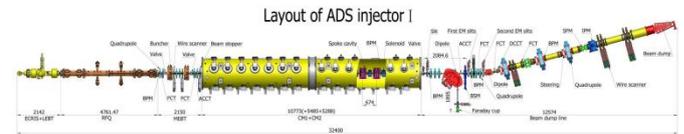

Figure 2: Schematic of C-ADS Injector I

### 2.1 LLRF HARDWARE ARCHITECTURE[2-3]

C-ADS injector-I uses mtca.4 platform and earliest mtca.4 digital board sis8300. This sis8300 board is equipped with a vertex 5 series FPGA. The vertex 5 series is strikingly lacks of logical resources, so only the main functions were developed during the verification process. Number of occupied slices reaches 99% after all logical implemented. The performances can be optimized with new FPGA chip in the future.

LLRF has an RF analog front-end, which generates sampling clocks and realize down conversion and up conversion functions. The sampling clock frequency is 104MHz, the intermediate frequency is 26MHz, and the LO frequency is 351MHz.

The RF system interlocking mostly includes surface temperature, water flow , water temperature , vacuum, multipactor current, ARC of SRF coupler,and also the ready signal from cryotron module. Once one of the interlocks is triggered, the RF will immediately shut down.

### 2.2 INTERFACE AND FUNCTIONS

The C-ADS INJ-I operates in continuous wave mode, but CSNS operates in pulse mode, so the system has been modified to pulse mode. Pulse feedforward control and rising edge control are used to reduce the burden of PI control.

\* Work supported by the Large Scientific Facility Open Subjectof Songshan Lake, Dongguan, Guangdong
† email address: xiezhexin@ihep.ac.cn

Some strategies for cavity protection have been implemented. Firstly, compare the error between the set point amplitude and the real-time cavity amplitude in the flat area of the pulse, and protect when the error exceeds the threshold. Secondly, the load Q factor is calculated for each pulse. If the load Q factor is below the threshold, the protection is triggered after the next pulse or several pulses.

Table 1: Parameter of SRF cavites for CSNS-II LINAC

| Items | SPOKE | SPOKE(Imp) | ELLIP |
|---|---|---|---|
| Freq(MHz) | 324 | 324 | 648 |
| Grad(Mv/m) | 7.3 | 9 | 14 |
| Length(m) | 0.694 | 0.694 | 0.86 |
| R/Q(Ω) | 410 | 410 | 309 |
| QL | 2.3E5 | 2.3E5 | 9.6E5 |
| LFD Hz/(MV/m)2 | -10.7 | -4.51 | -1.5 |
| f3db | 1408 | 1408 | 673 |
| df/dp(Hz/mbar) | 0.77 | 38 | 6.3 |

Table 2: Parameter of SRF LLRF for CSNS-II LINAC

| Items | SPOKE | UNIT |
|---|---|---|
| Freq(MHz) | 324 | MHz |
| RF Len | 7.3 | ms |
| Repetition | 25 | Hz |
| Beam Len | 500 | μs |
| Beam Cur | 30 | mA |
| Amp | ±0.3 | % |
| Ph | ±0.3 | deg |

The beam feedforward algorithm is used to compensate for beam interference. We have developed a sin/cos waveform to drive the piezoelectric controller to drive the piezoelectric in each pulse. The interface is shown in Fig.3.

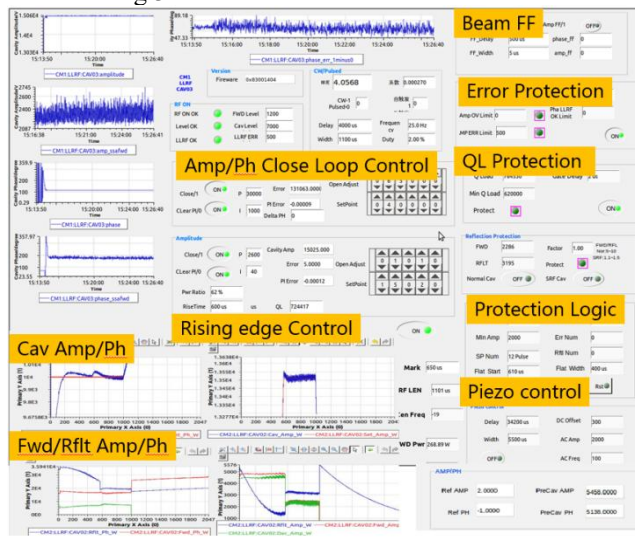

Figure 3: Interface of LLRF

### 2.3 Beam commission

The beam loading method is used to find the synchronous phase [4]. When the beam drifts through the unpowered cavity, the deceleration phase will be captured at the LLRF interface (Fig.4).

There is a certain degree of difference between the beam loading and phase scanning methods, which may be due to the pre detuning in the cavity during the experimental process.

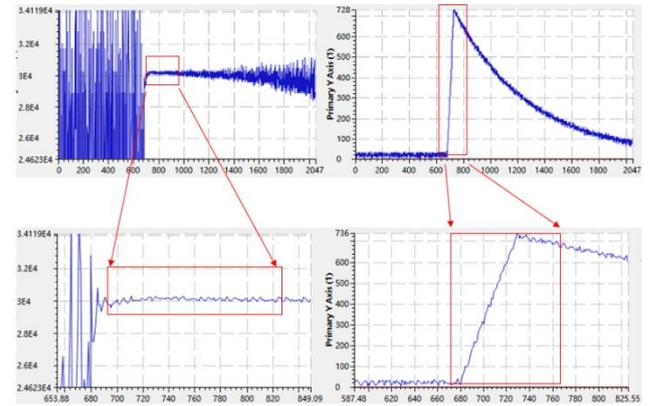

Figure 4: Beam loading: phase(Left) and amplitude(right)

A feedforward compensation method was used, and the vector method was used to obtain the feedforward vector to compensate for the beam current [5], achieving good results.

### 2.4 Experimental results

Ten SRF cavities operate stably for about 20 hours under 25 Hz 1.2 ms pulses, with amplitude and phase stability of about ± 0.3% and ± 0.3 °, meeting the requirements of CSNS-II. The use of new FPGA chips will achieve higher accuracy. The phase variation is shown in Fig.5.

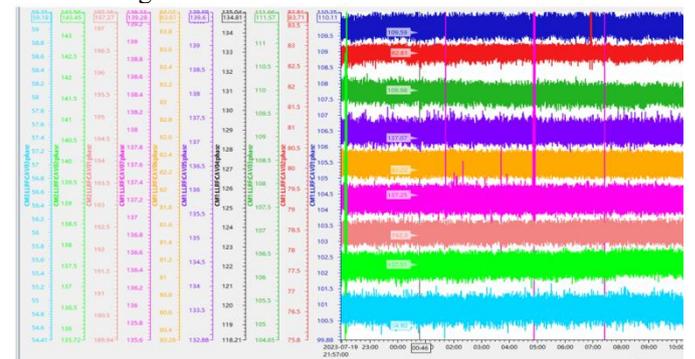

Figure 5: Long-term variation of the Cavity phase (about ±0.3°)

## CONCLUSION

The preliminary work of SRF LLRF is ongoing, the algorithm verification at C-ADS injector I shows a good result, the performance meeting the requirements of CSNS-II. In the future, new digital card will be used to pursue a higher goal. Automated program will developed to control about 50 SRF machines.